\documentclass[a4paper,USenglish]{article}

\usepackage[T1]{fontenc}
\usepackage[utf8]{inputenc}
\usepackage{authblk}
\usepackage{graphicx}
\usepackage{subfig}

\usepackage{color}
\usepackage{hyperref}
\hypersetup{colorlinks,
    citecolor=black,
    filecolor=black,
    linkcolor=black,
    urlcolor=black,
    plainpages=false}

\title{Automated Image Analysis of Hodgkin lymphoma}

\author[1]{Alexander Schmitz}
\author[1]{Tim Sch\"afer}
\author[1]{Hendrik Sch\"afer}
\author[2]{Claudia D\"oring}
\author[1]{J\"org Ackermann}
\author[1]{Norbert Dichter}
\author[2]{Sylvia Hartmann}
\author[2]{Martin-Leo Hansmann}
\author[1]{Ina Koch}
\affil[1]{Institute of Computer Science, Department of Molecular Bioinformatics\\
  Johann Wolfgang Goethe-University Frankfurt (Main), Robert-Mayer-Strasse 11--15, 60325 Frankfurt (Main), Germany\\
  \texttt{ina.koch@bioinformatik.uni-frankfurt.de}}
\affil[2]{Senckenberg Institute of Pathology\\
  Johann Wolfgang Goethe-University Frankfurt (Main), 60590 Frankfurt (Main), Germany\\
  \texttt{m.l.hansmann@em.uni-frankfurt.de}}

\begin{document}

\maketitle

\begin{abstract}
{\bf Motivation:}
Hodgkin lymphoma is an unusual type of lymphoma, arising from malignant B-cells. Morphological and immunohistochemical features of malignant cells and their distribution differ from other cancer types. Based on systematic tissue image analysis, computer-aided exploration can provide new insights into Hodgkin lymphoma pathology.

\noindent{\bf Results:}
In this paper, we report results from an image analysis of CD30 immunostained Hodgkin lymphoma tissue section images. To the best of our knowledge, this is the first systematic application of image analysis to a set of tissue sections of Hodgkin lymphoma. We have implemented an automatic procedure to handle and explore image data in Aperio's SVS format. We use pre-processing approaches on a down-scaled image to separate the image objects from the background. Then, we apply a supervised classification method to assign pixels to predefined classes. Our pre-processing method is able to separate the tissue content of images from the image background. We analyzed three immunohistologically defined groups, non-lymphoma and the two most common forms of Hodgkin lymphoma, nodular sclerosis and mixed cellularity type. We found that nodular sclerosis and non-lymphoma images exhibit different amounts of CD30 stain, whereas mixed cellularity type exhibits a large variance and overlaps with the other groups. The results can be seen as a first step to computationally identify tumor regions in the images. This allows us to focus on these regions when performing  computationally expensive tasks like object detection in the high-resolution image.

{\bf Contact:}
\href{ina.koch@bioinformatik.uni-frankfurt.de}{ina.koch@bioinformatik.uni-frankfurt.de}, \href{m.l.hansmann@em.uni-frankfurt.de} {m.l.hansmann@em.uni-frankfurt.de}
\end{abstract}

\section{Introduction}
In this paper, we describe the analysis of tissue image data of Hodgkin lymphoma using computational methods.

\subsection{Hodgkin lymphoma}
\label{sec:hl}
Hodgkin lymphoma ({\sc hl}) is a type of lymphoma, originating from B-cells in most of the cases~\cite{Kueppers1994}. In {\sc hl}, neoplastic cells circumvent the immune surveillance and apoptotic control. A key characteristic in {\sc hl} is that the malignant cells make up only about 1\% of the tumor mass, while being outnumbered by a microenvironment of reactive lymphoid cells~\cite{Steidl2011, Mathas2009}. Depending on the infiltrate, morphological, phenotypic, genotypic, and clinical findings, {\sc hl} is further subdivided. According to the WHO classification, the main subtypes are classical  {\sc hl} (c{\sc hl}) and nodular lymphocyte-predominant {\sc hl} ({\sc nlp}~{\sc hl}).

\begin{figure}[!ht]
\centering
\includegraphics[width={0.7\textwidth}]{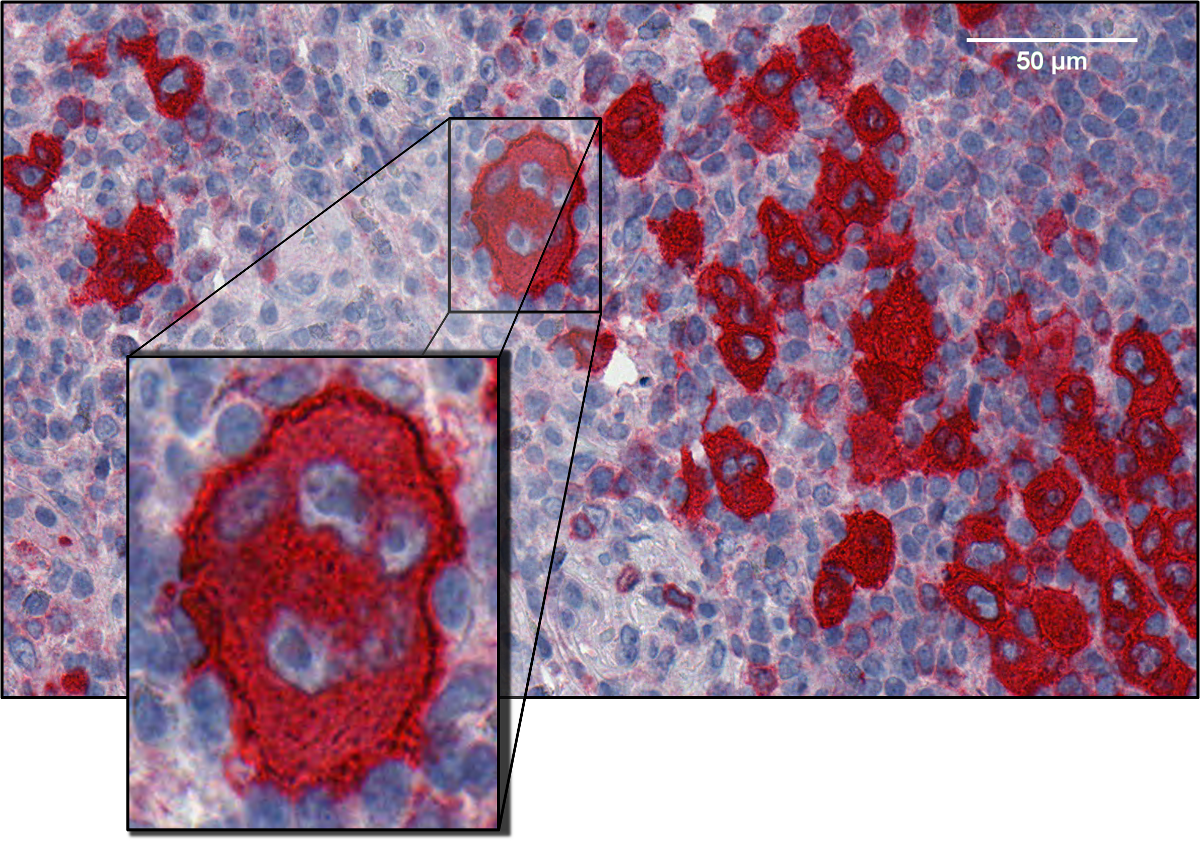}
\caption[$CD30^+$ Reed-Sternberg cell]{\textbf{CD30$^+$ Reed-Sternberg cell.} Malignant cells are marked with CD30 immunostain. Additional hematoxylin counterstain is provided to visualize nuclei. A giant {\sc rs} cell can be seen in the center of the image, outnumbered by a variety of other lymphoid cells (nuclei are visible). Multiple irregularly shaped nuclei are visible, a distinguishing feature of this cell.}
\label{fig:rs_cell}
\end{figure}

The assignment to c{\sc hl} is based on the identification of characteristic tumor cells. These cells are termed Hodgkin and Reed-Sternberg cells ({\sc hrs} cells)~\cite{Reed1902, Sternberg1898}. For immunophenotypic differentiation of {\sc hl}, methods from the field of immunohistochemistry are used. Of particular interest for this work are the cluster-of-differentiation protocols (CD), used to identify cell surface proteins presented on leukocytes (e.g., CD30, CD20,  CD15). CD30 belongs to the tumor necrosis receptor family (TNFR) and is usually expressed by {\sc hrs} cells. Figure \ref{fig:rs_cell} illustrates a typical CD30$^+$ immunostained {\sc hrs} cell within its microenvironment. Hematoxylin counterstaining is applied to visualize cell nuclei. c{\sc hl} is further divided into subtypes depending on immunophenotypic findings and the composition of the cellular microenvironment; the most common subtypes are termed nodular sclerosis ({\sc ns} c{\sc hl}), and mixed cellularity ({\sc mc} c{\sc hl})~\cite{Hansmann2002}. All subtypes of c{\sc hl} share an immunophenotype of CD30$^+$, CD15$^+$, and CD20$^-$~\cite{Kueppers2009a}.

Moreover, the expression of various cytokines and chemokines is thought to be responsible for the attraction of different cells, such as small lymphocytes, macrophages, mast cells, plasma cells, stromal cells, histiocytes, or fibroblasts~\cite{Aldinucci2010, Steidl2011}. The extent of cellular infiltration, involving many different cell types as a response to inflammatory signals, represents a key characteristic of {\sc hl}. The events that happen in the course of malignancy development as well as the complex interaction network of malignant cells in {\sc hl} remain to be resolved. This can be done using image processing, which may provide new insights into the pathology of {\sc hl}. 

\subsection{Whole Slide Image Analysis}
There is a broad range of literature on image analysis of non-lymphoid tumors. Automated methods for the analysis of histology images have been employed in various cancer types to detect, classify, and quantify cells. A common issue is the segmentation of cell nuclei. For instance, Naik \textit{et al.} used a Bayesian nuclear segmentation scheme in prostate and breast cancer histopathology, integrating color, texture, and shape information~\cite{Naik2008}. Bayesian models with additional contextual information from Markov random fields have been applied as a probabilistic approach for prostate cancer detection~\cite{Monaco2009}. More general approaches for hematoxylin \& eosin (H\&E) stained images have also been applied. Lei \textit{et al.} used Gaussian mixture models to perform local and global clustering, extracting different tissue constituents in cervix histology images~\cite{Lei2011}. Sertel \textit{et al.} developed a system for segmentation of eosinophilic and basophilic structures in H\&E stained tissue section images of neuroblastoma. Different tissue subtypes were classified using texture features and a new set of structural features~\cite{Sertel2009}.

Large-scale computations and machine learning algorithms have been applied to breast histology images~\cite{Petushi2006}. A hybrid segmentation approach and supervised classification scheme hve been used to identify micro-textures meaningful for differentiation of tissue types. Karacali \textit{et al.} proposed a high-throughput method of texture heterogeneity on breast tissue images to identify regions of interest. They subdivide the images into small image blocks and pass these blocks to a texture-based statistical learning algorithm, which classifies them into categories of normal, malignant, and non-specific tissue~\cite{Karacali2007a}.

Some investigations were performed on non-Hodgkin lymphoma. An automatic classification of three types of malignant lymphoma has been proposed by Orlov \textit{et al.}, taking into account chronic lymphocytic leukemia, follicular lymphoma ({\sc fl}), and mantle cell lymphoma~\cite{Orlov2010}. The classification is based on a supervised classifier (WND-CHARM), applying additional feature weights learned in the classifier's training phase. They achieved high accuracy in discriminating the different lymphoma types using various statistical and texture features. {\sc fl} arises from the malignancy of follicle center B cells. Sertel \textit{et al.} developed a computerized system to detect follicles based on texture features in an immunostained image. This is followed by the detection of centroblasts within the follicular regions~\cite{Sertel2008b}.

Today, a number of software solutions for analyzing images is available. Commercial software includes general tools like MATLAB~\cite{Matlab} as well as software geared especially towards digital pathology like Aperio ImageScope~\cite{Aperio}, Definiens Tissue Studio~\cite{Definiens}, and many other. There exist also freely available tools, the most prominent of which are CellProfiler~\cite{Lamprecht2007}, ImageJ~\cite{Abramoff2004}, GNU Octave~\cite{Eaton2008}, and Scilab~\cite{Scilab2012}.

Despite all these approaches, no work exists for {\sc hl}. {\sc ns}~c{\sc hl} (nodular sclerosis) and {\sc mc}~c{\sc hl} (mixed cellularity) represent two different types of {\sc hl} which often have distinct patterns of tumor cell distribution. In {\sc ns} c{\sc hl}, the tumor cells grow in a nodular pattern with broad collagen bands separating the nodules. In contrast, {\sc mc} c{\sc hl} tumor cells are more isolated and often distributed over the whole tissue section. {\sc ns} c{\sc hl} accounts for about 64\%, whereas {\sc mc} c{\sc hl} is found in about 30\% of the c{\sc hl} cases. While the visual differentiation between the two types of c{\sc hl} is possible in many cases, non-lymphoma images sometimes exhibit features similar to c{\sc hl} subtypes. An example is lymphadenitis, an infection of the lymph nodes, which may result from certain bacterial infections and often leads to inflammation and swelling of lymph nodes. 

Our goal is to analyze the considered types based on CD30 stained tissue sections, using pixel based image processing techniques. We use our own implementation to be flexible in handling and exploring {\sc hl} whole slide tissue-images and to easily extend the software in future projects. Our software is intended to evolve into a tool that extracts statistical information from a large database of Hodgkin lymphoma images. It will be combined with existing imaging software solutions like CellProfiler~\cite{Lamprecht2007}.

\section{Material}
The source images used in this study are CD30 stained tissue slides of lymph node sections. The tissue sections have been pretreated, and immunostainings for CD30 were performed as described previously~\cite{Hartmann2011}. Hematoxylin counterstain was applied to stain cell nuclei. The slides were assigned by pathologists to one of the following groups:

\begin{itemize}
\item Nodular sclerosis classical Hodgkin lymphoma ({\sc ns}~c{\sc hl})
\item Mixed cellularity classical Hodgkin lymphoma ({\sc mc}~c{\sc hl})
\item Non-lymphoma (lymphadenitis with and without follicular hyperplasia)
\end{itemize}

All samples were randomly taken from an anonymized data set from the \emph{Senckenberg Institute of Pathology Frankfurt am Main}. Tissue sections of small size may not be repre\-sentative for the complete lymph node and were discarded. Furthermore, only tissue sections with distinct immunostaining were taken into account. In these images, CD30$^+$ cells can be visually distinguished from the background. This selection has been carried out by pathologists yielding 62 images of {\sc ns} c{\sc hl}, 57 of {\sc mc} c{\sc hl}, and 51 of non-lymphoma.

The digital slides were captured using an \emph{Aperio ScanScope XT} scanning device with a 40x objective lens. The typical size of the tissue samples is about 15 mm$^2$, and the resulting images reach dimensions of up to 100,000 x 100,000 pixels (0.25$\mu m$ per pixel). The images are provided in the Aperio SVS format, a single-file pyramidal TIFF with non-standard metadata and compression. Our input files contain four levels with different resolutions of the original image, each of which is split into tiles. For our analysis, we use the second level which is down-scaled by a factor of 16 on each axis compared to the full-resolution image.

\section{Methods}
We use Openslide 3.2 to handle whole slide images and implement our analysis applying the Java Advanced Imaging API~\cite{Goode2008}. The input images contain stained tissue in front of a bright background. They may also include artifacts like air bubbles, small tissue fragments and stain residues. To eliminate such artifacts, as well as background, we apply image pre-processing steps. 

\subsection{Pre-Processing}
\label{sec:pre}
During pre-processing, we use Gaussian filtering, thresholding, and region labeling to identify the tissue region as described in~\cite{Gonzalez2007}. We analyze the tissue region in the following classification steps.  

\paragraph{Gaussian filtering}
We convolve the input image with a Gaussian filter which is applied to all three RGB channels. This step blurs the input image by replacing the RGB values of each pixel by the weighted RGB values of a 43 $\times$ 43 window. The weights follow a Gaussian distribution with a variance of $\sigma = 3$ pixels. The images are prepared for the threshold process by filtering. Thereby, pixels inside the tissue area are kept from falling into the background category. 

\paragraph{Threshold background pixels} 
The usage of a threshold becomes feasible since all images share a distinct background peak in the histogram. We convert the images to gray scale by averaging the RGB values of the three channels and then split the histogram of the resulting image into two distinct classes by choosing a threshold value. Thus, pixels are labeled as either object or background pixels during this process. A threshold is applied to the smoothed image using the brightness value of the local minimum next to the background peak. Pixels labeled as background are removed from the original image whereas those of tissue regions are retained. 

\paragraph{Filtering tissue regions}
The remaining pixels can be separated into connected areas using a region labeling method. Connected areas correspond to pixel regions, such that a connecting path exists between each pair of pixels. The region labeling algorithm identifies all connected regions within the image. Only pixels belonging to a sufficiently large area are retained. We choose a region size of $40,000$ pixels as threshold. This results in discarding small separated tissue fragments and spots on the object slide. Figure \ref{fig:prepro} demonstrates an example for the effects of these three pre-processing steps.

\subsection{ Pixel Classification }
\label{sec:class_quant}
We use the following terminology: all images assigned by pathologists to the same type ({\sc ns}~c{\sc hl}, {\sc mc}~c{\sc hl} or {\sc nl}) are called a \emph{group}. During pixel classification, pixels are assigned to one of the \emph{pixel classes} defined in this section.

After pre-processing, the pixels considered as tissue are classified using a supervised approach. For each pixel, a set of descriptors is computed as signature. We apply a minimum distance classifier, which is trained on examples for each pixel class, resulting in one mean signature, a vector of descriptor values. Therefore, we manually prepared a set of 36 image sections which represent the pixel classes (6 images per class, image size between 150 and 1000 pixels). The classifier assigns a pixel to the reference pixel class whose signature is closest to the pixel's signature. To compute the distance between pixel class and a sample signature a predefined Euclidean distance measure is applied. The mean signatures for the following pixel classes are determined:

\begin{itemize}
\item \textbf{Background}: The pre-processing does not eliminate all pixels that may be assigned to background. Therefore, the class  \emph{Background} represents such pixels.

\item \textbf{Low intensity}:
Pixels having very low intensity should also be regarded as non-tissue.

\item \textbf{Hematoxylin$^+$}:
Cell nuclei are hematoxylin positive and stained in deep blue.

\item \textbf{CD30$^+$}:
CD30$^+$ stained regions are typical for {\sc rs} cells. Thus, CD30 is an indicator for possible tumor cells. 

\item \textbf{Nonspecific red}:
These are image regions which are colored red, but not due to CD30 staining. Examples include prolate regions resembling vessels.

\item \textbf{Unstained}:
Nonspecifically stained regions are slightly colored.
\end{itemize}

To compute the signature we use a set of low-level pixel-based descriptors. All descriptors are solely based on the pixel intensity values. The set of descriptors is defined as follows:

\begin{itemize}
\item \textbf{Brightness Descriptor}:
The mean value over all three color channels red (R), green (G), and blue (B) of a considered pixel is computed. 

\item \textbf{Mean Descriptor}:
For all pixels within a distance of one pixel using an 8-connected neighborhood, we take the average values of the R, G, and B channels. 

\item \textbf{CD30 Saturation}:
To account for the CD30 staining, we use a modified saturation descriptor, which is given in Equation \ref{eq:sat}. We only consider two stainings which are red and blue, so green pixels can only be produced by artifacts or noise.

\begin{equation}
\label{eq:sat}
S_{CD30}=(max(|R-B|,|R-G|)-min(|R-B|,|R-G|,|B-G|))
\end{equation}
\end{itemize}

Finally, we compute the number of pixels which have been assigned to the defined pixel classes for each image. This results in a vector for each image containing the pixel counts for each pixel class. The counts of \textit{Unstained}, \textit{CD30$^+$}, \textit{Nonspecific red}, and \textit{Hematoxylin$^+$} represent the tissue classes. Since we are interested in the tissue classes, we omit the counts of \textit{Background} and \textit{Low intensity}. For the tissue classes, we compute the fraction of total pixels belonging to the respective class.

\section{Results and Discussion}
\begin{figure}[!ht]
\centering
\includegraphics[width={1.0\textwidth}]{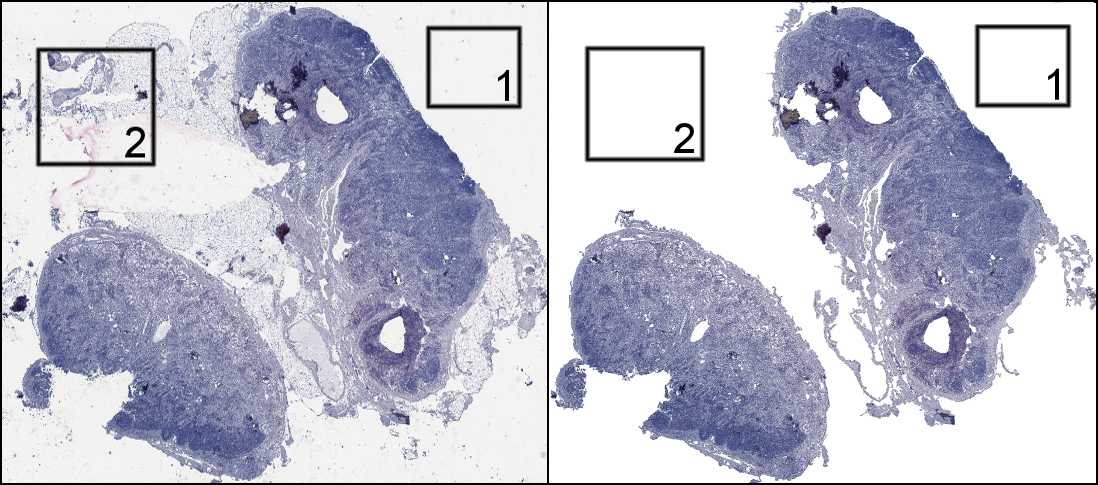}
\caption[Pre-processing steps applied to example tissue slide]{\textbf{Pre-processing steps applied to example tissue slide.}
Comparison of tissue section before (left) and after (right) the pre-processing step. The image is first processed using Gaussian filtering. Background thresholding, followed by a region labeling algorithm is then applied to remove fragments and small particles, which may occur during the preparation process. In Box 1, the removal of background pixels is illustrated. Box 2 depicts the elimination of tissue fragments and connective tissue.}
\label{fig:prepro}
\end{figure}

The pre-processing steps are able to identify regions of interest and discard background information. Figure \ref{fig:prepro} depicts an example for the effects of the pre-processing. As visible in Box 1, background pixels are rejected. Furthermore, tissue fragments and connective tissue areas are also ignored (Box 2).

\begin{figure}[!htbp]
\centering
\includegraphics[width={\textwidth}]{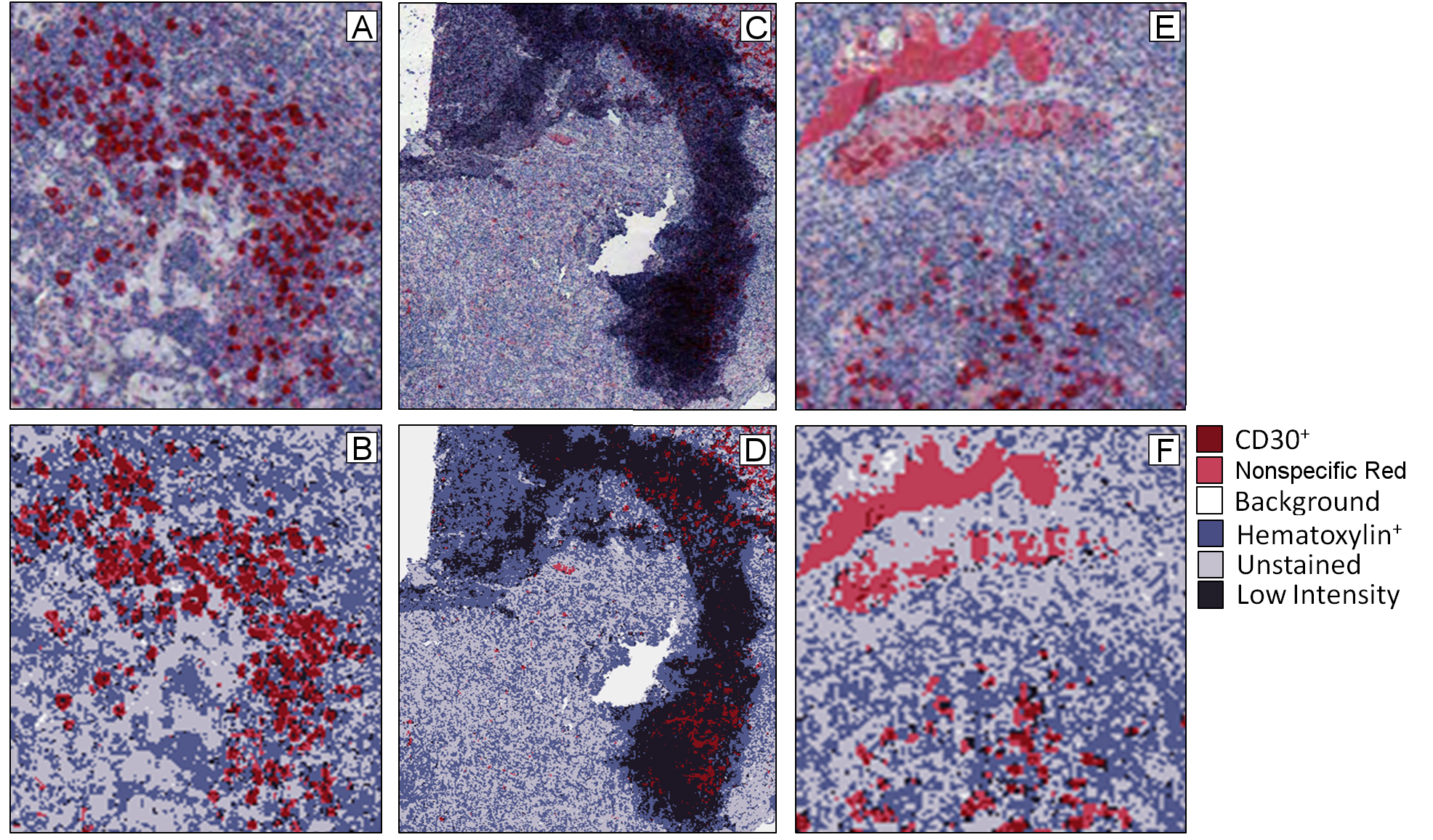}
\caption[Detailed view of classified image sections.]{Detailed view of classified image sections. Images A and B depict the classification results for \emph{CD30$^+$} regions. In C and D, the classification results for areas with tissue folds are depicted. E and F present the distinction between \emph{CD30$^+$} and \emph{Nonspecific red}.}
\label{fig:classification_detail}
\end{figure}

The pixel classes referred to in this paper reflect one way of describing the content of histological images. These classes were found to give reasonable results for the underlying image data. For a more detailed view, Figure \ref{fig:classification_detail} depicts the results of pixel-based classification for small image regions. On the top, the original image region is shown, whereas on the bottom the resulting image is depicted. Images A and B demonstrate that \emph{CD30$^+$} regions are correctly assigned in most of the cases. Moreover, \emph{Unstained} and \emph{Hematoxylin$^+$} regions are separated from each other. The class \emph{Low intensity} is introduced to primarily cover areas, where tissue folds and disruptions occur. In such areas, it is difficult to assign pixels to tissue classes. Therefore, we want to exclude these areas altogether. For instance, Figure \ref{fig:classification_detail}C contains an area of disrupted tissue. Pixels within this region have low intensity values. As visible in Figure \ref{fig:classification_detail}D, the classifier assigns most of these pixels to the class \emph{Low intensity}. In Figure \ref{fig:classification_detail}E, a red area without distinct CD30 staining is visible. Pixels within this region are mostly assigned to \emph{Nonspecific red}. From these figures it can be inferred that the pixel-based classification is able to identify the supplied training classes based solely on color information. This could be also confirmed by visual examination of the resulting images.

\begin{figure}[!ht]
   \begin{center}
   \subfloat[Section of an {\sc ns} c{\sc hl} example image.] {\label{fig:orig_img}\includegraphics[width={\textwidth}]{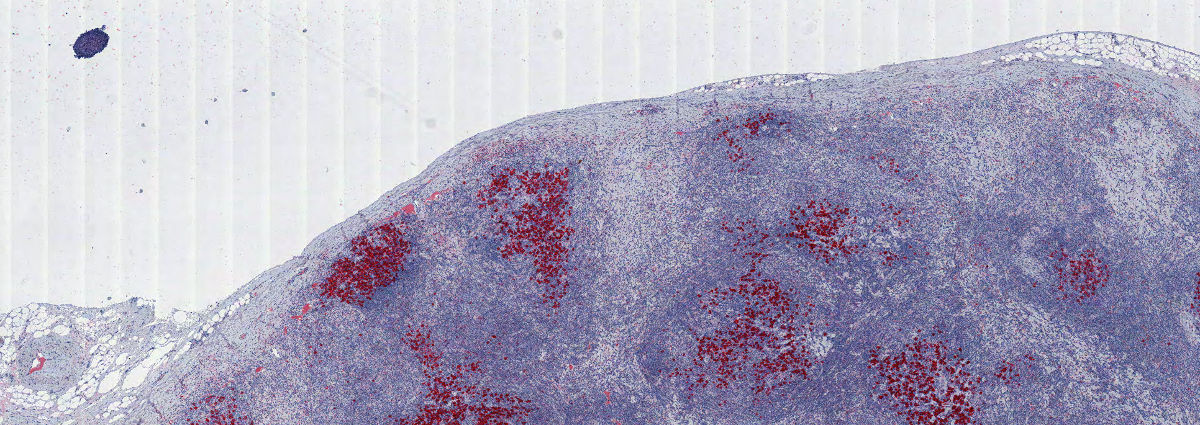}}\\
   \subfloat[Region of interest based on the pixel-classified image.]{\label{fig:roi_img}\includegraphics[width={\textwidth}]{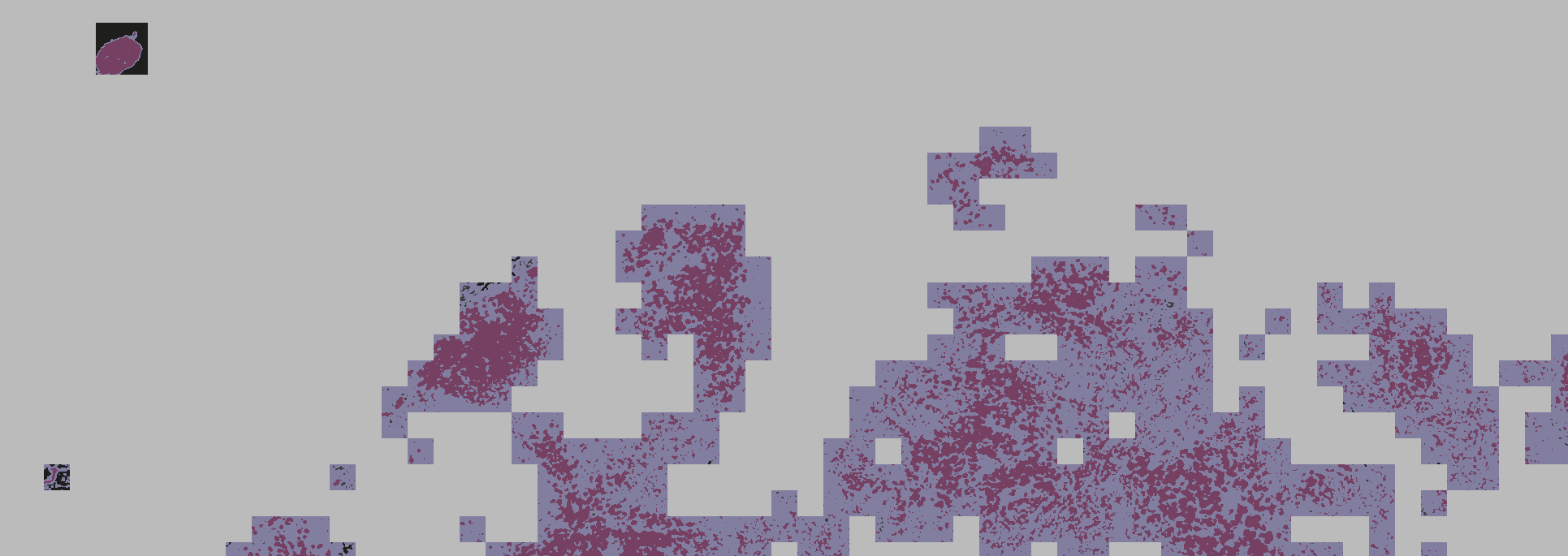}}\\
   \caption{Results of region of interest identification in whole slide images. The pixel-classified image is reduced to the region of interest containing CD30$^+$ areas. Only tiles of the original image which contain possible CD30$^+$ cells are kept, tiles without CD30$^+$ regions are gray.}
   \label{fig:roi}
   \end{center}
\end{figure}

The results of the pixel classification allow us to identify image regions containing CD30$^+$ cells. As CD30 is a marker for {\sc hrs} cells, information on these sections can be used to quickly select regions in the full-resolution level of the images, where candidates for {\sc hrs} cells exist. To filter the images for these regions of interest, we apply a grid which subdivides the image into tiles of size 32 $\times$ 32 pixels, which corresponds to an area of 512 $\times$ 512 pixels in the full-resolution image. Then, we discard all tiles which do not contain any \emph{CD30$^+$} pixel, see Figure \ref{fig:roi}. In Figure \ref{fig:roi}B, we see that a large amount of tiles does not contain candidate CD30$^+$ cells. For the whole image, about 75\% of the tiles could be ignored. This substantially reduces the computational costs of further processing steps, e.g., object detection in the full-resolution image.

\begin{figure}[!ht]
   \begin{center}
   \subfloat[Quantification results for non-lymphoma tissue sections.] {\label{fig:res_nl}\includegraphics[width={0.6\textwidth}]{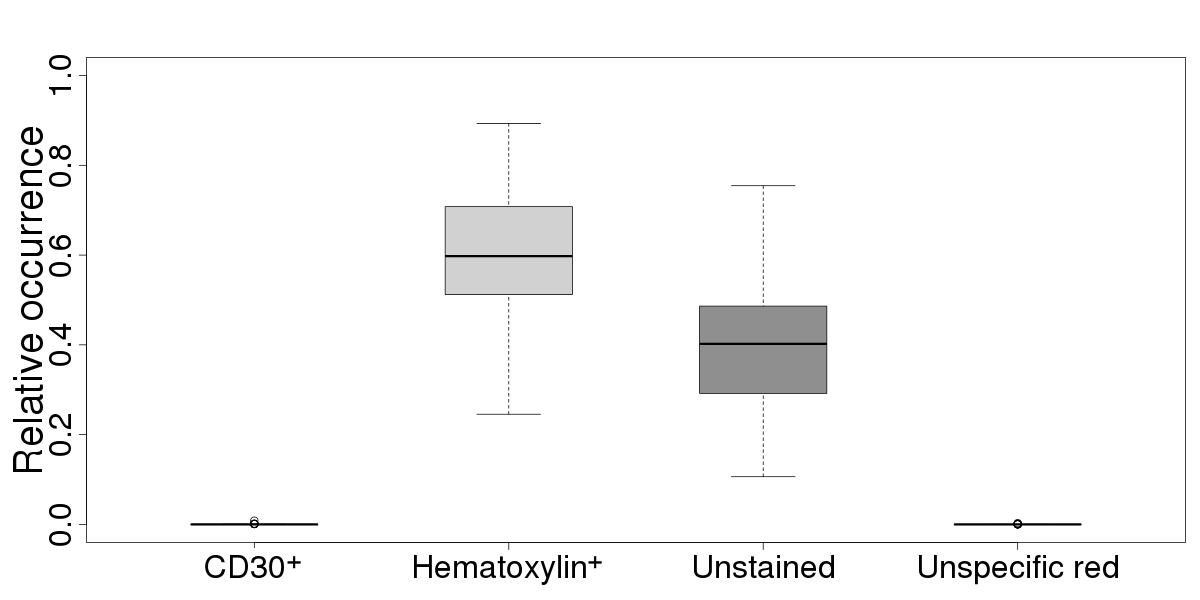}}\\
   \subfloat[Quantification results for {\sc ns} c{\sc hl} tissue sections.]{\label{fig:res_ns}\includegraphics[width={0.6\textwidth}]{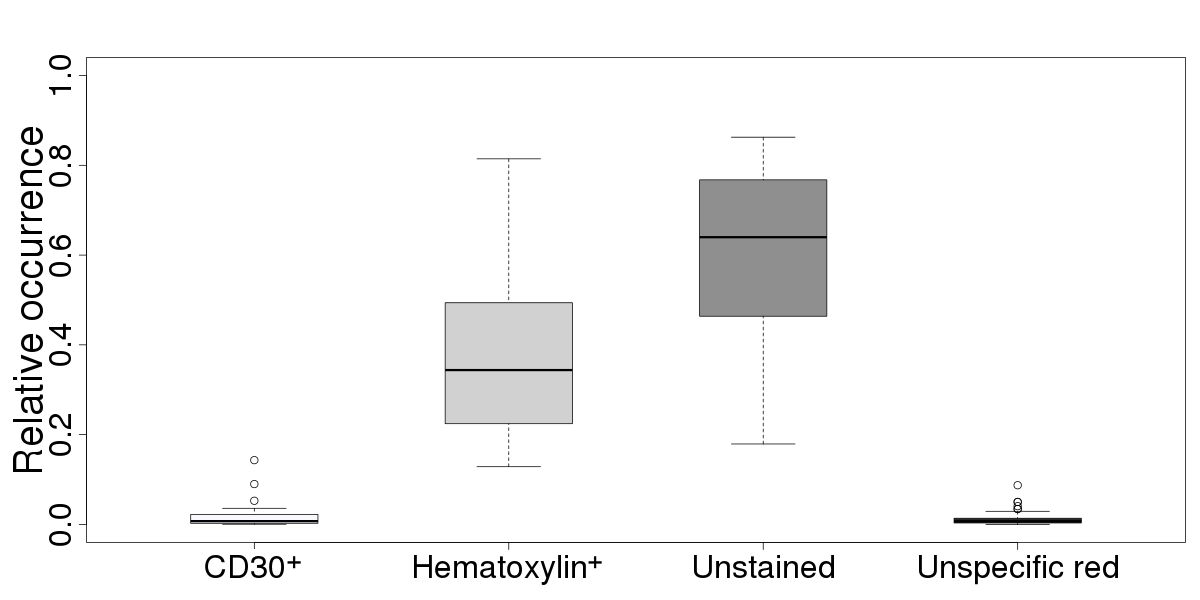}}\\
   \subfloat[Quantification results for {\sc mc} c{\sc hl} tissue sections.] {\label{fig:res_mc}\includegraphics[width={0.6\textwidth}]{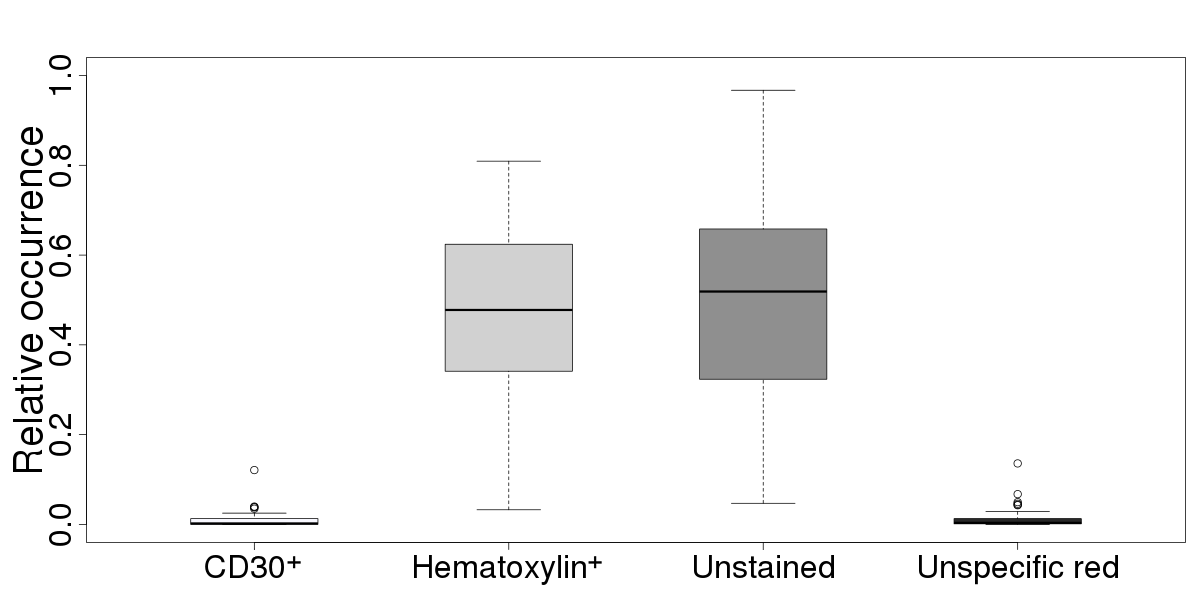}}
   \caption[Quantification results for non-lymphoma, {\sc ns} c{\sc hl}, and {\sc mc} c{\sc hl} tissue sections.]{\textbf{Quantification results for non-lymphoma, {\sc ns}~c{\sc hl}, and {\sc mc}~c{\sc hl} tissue sections.} The results present the relative amount of pixels which have been assigned to the tissue classes. \textit{CD30$^+$}, \textit{Hematoxylin$^+$}, \textit{Nonspecific red}, and \textit{Unstained} are considered as belonging to tissue. Pixels assigned to either the class \textit{Background} or \textit{Low Intensity} are taken together as non-tissue (not shown). For each class, the frequencies are averaged over all considered cases of non-lymphoma, {\sc ns} c{\sc hl}, and {\sc mc} c{\sc hl}, respectively.}
   \label{fig:res_all}
   \end{center}
\end{figure}

The quantification results for non-lymphoma, {\sc ns} c{\sc hl}, and {\sc mc} c{\sc hl} are depicted in Figure~\ref{fig:res_all}. The relative amount of CD30$^+$ pixels in non-lymphoma cases is considerably lower than in {\sc ns}~c{\sc hl} and {\sc mc}~c{\sc hl}, see Figure~\ref{fig:res_all}A. This is expected because CD30$^+$ Hodgkin and {\sc rs} cells are a characteristic feature of c{\sc hl}. However, small amounts of CD30$^+$ cells are known to occur in non-lymphoma cases and can be visually confirmed in the source images.

The amount of \emph{CD30$^+$} in the groups is proportional to the amount of \emph{Nonspecific red}, due to the fact that the signatures of these groups are very similar. We use down-scaled versions of the original images, so a single pixel represents several pixels of the full-resolution image. Therefore, pixels at the borders of \emph{CD30$^+$} regions are averaged with their neighbors and may thus be classified as \emph{Nonspecific red}. Nevertheless, visual inspection of the resulting images suggests that most of the red-colored regions without CD30 staining are correctly assigned to this pixel class.  

The number of pixels classified as \emph{Hematoxylin$^+$} in tissue regions is rather low in {\sc ns}~c{\sc hl} (compare Figure~\ref{fig:res_all}B). \emph{Hematoxylin$^+$} can be understood as a label for regions with a high density of cell nuclei. The low number may correspond to the high amount of fibrotic bands occurring in this type of c{\sc hl}. Figure~\ref{fig:res_all}C illustrates that the distributions for the pixel classes of {\sc mc}~c{\sc hl} have a clear overlap with both other groups. 

In conclusion, this suggests that all three groups contain images which are hard to classify by only considering their \emph{CD30$^+$} share. Due to the large overlap with both other groups, we assume that {\sc mc}~c{\sc hl} is hard to distinguish from {\sc nl}, and {\sc ns}~c{\sc hl}. This assumption has been confirmed by an image classification approach using a minimum distance to mean classifier~\cite{Gonzalez2007}. The classifier is based on the amount of \emph{CD30$^+$} pixels only. 50.6\% of the images were classified correctly ({\sc ns}~c{\sc hl}: 34/62, {\sc mc}~c{\sc hl}: 26/57, {\sc nl}: 26/51). This is an enrichment of 51.8\% compared to a random draw, which would yield 33.33\% correct assignments. It becomes apparent that pixel-based classification does not stand up to the current state-of-the-art approaches in image classification, as more sophisticated methods are reported to reach correct classification rates of up to 96\% for similar images~\cite{Gurcan2009}.

\section{Conclusion}
In this work, we presented results of our automated image analysis applied to lymph node image data. To the best of our knowledge, there does not yet exist a systematic application of image analysis to {\sc hl}. Standard pre-processing methods like Gaussian filtering, application of a threshold for background elimination and region labeling for identification of relevant tissue patches, were applied.

After pre-processing, a supervised classification was used to assign each pixel to one of the pixel classes: \emph{Background}, \emph{Low intensity}, \emph{Unstained}, \emph{CD30$^+$}, \emph{Hematoxylin$^+$}, and \emph{Nonspecific red}. For the quantification, we consider only tissue classes, i.e., pixels classified as \emph{Unstained}, \emph{CD30$^+$}, \emph{Hematoxylin$^+$}, or \emph{Nonspecific red}. For each pixel class, we computed the relative fraction of pixels belonging to the respective class. 

We observed a large variation concerning the occurrences of pixel classes in the groups. This variation could be related to the large amount of images in each group and the fact that various stages of {\sc hl} and different types of non-lymphoma cases were investigated. In advanced stages of {\sc hl} disease, the amount of CD30$^+$ cells increases whereas in early stages only few malignant cells can be detected. This may lead to statistical outliers within the groups. 

Visual inspection of the images indicates that the spatial distribution of CD30$^+$ cells differs between the image groups. For instance, in {\sc ns}~c{\sc hl} the CD30$^+$ cells are arranged in a nodular pattern whereas in most cases of {\sc mc}~c{\sc hl} or {\sc nl} CD30$^+$ cells seem to be equally distributed over the entire tissue section. Therefore, we conclude that the usage of descriptors that include spatial information and additional tissue classes will improve the classification results. We intend to apply such descriptors using a combination of our software and existing implementations, like CellProfiler. 

Because malignant cells in {\sc hl} contribute only a small proportion of about 1\% of the total cell count, large regions exist in the images which do not contain any {\sc hrs} cell. The tile-based filtering approach addressed in Figure \ref{fig:roi} may thus contribute to reduce computational cost of subsequent processing steps. In the near future, we plan to extend our approach by further analysis methods to gain high-level information from the image data which can then be used to address questions in systems biology.

\bibliography{../bib/biball} 

\end{document}